\documentstyle[preprint,aps]{revtex}
\tightenlines
\def\Journal#1#2#3#4{{#1} {\bf #2}, #3 (#4)}

\def\CMP{{\em Commun. Math. Phys.}}
\def\IJMPA{{\em Int. J. Mod. Phys.} A}
\def\IJMPB{{\em Int. J. Mod. Phys.} B}
\def\MPLA{{\em Mod. Phys. Lett.} A}

\def\NPB{{\em Nucl. Phys.} B}

\def\PLB{{\em Phys. Lett.} B}

\def\PRA{{\em Phys. Rev.} A}

\def\PRD{{\em Phys. Rev.} D}

\def\PREPC{{\em Phys. Rep.} C}
\def\RMP{{\em Rev. Mod. Phys.}}

\def\ZPC{{\em Z. Phys.} C}

\newcommand{\be}{\begin{equation}}
\newcommand{\ee}{\end{equation}}
\newcommand{\bea}{\begin{eqnarray}}
\newcommand{\eea}{\end{eqnarray}}

\newcommand{\hf} {{1\over2}}
\newcommand{\nonu}{\nonumber\\}
\def\dk{\Delta k}
\def\tr{{tr}}
\def\la{\lambda}
\def\eq#1{(\ref{#1})}
\def\cm{{\cal M}}
\def\ord{{\cal O}}

\begin{document}
\title{Renormalization group for the internal space}

\author{Jean Alexandre\thanks{alexandr@lpt1.u-strasbg.fr}}
\address{Laboratory of Theoretical Physics, Louis Pasteur University\\
3 rue de l'Universit\'e 67087 Strasbourg, Cedex, France}

\author{Janos Polonyi\thanks{polonyi@fresnel.u-strasbg.fr}}
\address{Laboratory of Theoretical Physics, Louis Pasteur University\\
3 rue de l'Universit\'e 67087 Strasbourg, Cedex, France\\
and\\
Department of Atomic Physics, L. E\"otv\"os University\\
P\'azm\'any P. S\'et\'any 1/A 1117 Budapest, Hungary}
\date{\today}
\maketitle
\begin{abstract}
The renormalization group method is a successive integration over the
fluctuations which are ordered according to their length scale, a parameter in 
the
external space. A different procedure is described, where the fluctuations
are treated in a successive manner, as well, but their order is given by
an internal space parameter, their amplitude. The differential version
of the renormalization group equation is given which is the functional
generalization of the Callan-Symanzik equation in one special case and 
resums the loop expansion in another one.
\end{abstract}

\vfil
\eject

\section{Introduction}
The renormalization group method is usually applied in two
different manners. The original way is to provide 
an insight into the scale dependence
of the coupling constants \cite{wilsrg}. Another, more recent use
is to perform a partial resummation of the perturbation 
expansion by making an infinitesimal change of the cutoff 
in a time and using the functional formalism \cite{wh}-\cite{tetra}.
Both goals are realized by the blocking procedure, the 
successive elimination of the degrees of freedom 
which lie above the running ultraviolet cutoff. The resulting
evolution equation yields the dependence of the coupling 
constants in the cutoff which is a scale parameter introduced
in the space-time, the external space. 
A difficult problem in this program is the lack of 
explict gauge invariance. Since the separation of the modes
according to their length scales is not gauge invariant,
the gauge symmetry is unavoidably lost during the blocking.

We propose here a different slicing procedure of the
path integral, the gradual increase of the
amplitude of the fluctuations. We introduce a cutoff
in the internal space\footnote{The internal and the external 
spaces stand for the space of the field amplitude 
and the space-time manifold, respectively.}
by constraining the amplitude
of the fluctuating field and obtain the evolution equation
by the gradual release of this constraint. 
Since the field variable is usually dimensional
this procedure can be interpreted in terms of a scale dependence,
as well. More precisely, the mass parameter of the lagrangian
serves as a cutoff for both spaces thereby providing
a dynamical correspondence between the scales in the internal and
the external spaces.
An important feature of this scheme is that the 
scale parameter, the mass in particular, is introduced by hand in the 
internal space. The length scale of the external space is, on the contrary,
induced dynamically. Such a dynamically generated length scale in the
external space appears more natural and avoids 
the problems mentioned above. In fact, having
imposed the cutoff in the internal space the gradient
expansion, being an approximation procedure in the external space,
goes through without problem. The gauge invariance
can be maintained, as well, the only requirement is
that the suppression of the fluctuations must be
achieved in a gauge independent fashion, a condition
easy to comply with in the internal space. This latter improvement
is somehow reminiscent of the achievement of the Higgs mechanism 
where a dynamically generated mass is produced
without spoiling the gauge symmetry.

There are further advantages in making in the internal space 
the analogue procedure of the blocking.
One is to get a new insight
into the origin of the anomalous dimension.
To see this we recall that the internal space has 
actually been used for the purpose of 
establishing the renormalization group. The renormalization
conditions was imposed at a non-vanishing background field
field in ref. \cite{cowe} providing the scale for the
internal space. The resulting renormalization group function was computed
in the one-loop level and found to be identical with
the usual functions, resulting from the renormalization
conditions introduced in the external space. This
agreement between the scale dependence in the external 
and internal space is not generally present and the two-loops
results should already yield a difference. In fact,
the scales inferred from these two spaces are related
by the terms of the lagrangian which mix the internal and the
external spaces, the terms with space-time derivatives.
The simplest of them is the wave function renormalization
constant, $\phi_B(x)=Z^{1/2}\phi_R(x)$,
which is trivial for a scalar model in the one-loop approximation. 
The anomalous dimension is nothing but the demonstration
of the difference between the scale dependence in the 
internal and the external spaces.

Another bonus of the internal space renormalization group
is the generalization of the Callan-Symanzik equation
for the functional formalism. The Callan-Symanzik
equation describes the dependence of the
Green functions on the mass \cite{casy}. Since the mass parameter
controls the width of the peak in the wave functional around
the vacuum configuration the mass can be thought as the
simplest cutoff parameter in the internal space.
The similarity of the renormalization group coefficient functions
obtained by the bona fide renormalization group and the
Callan-Symanzik equation demonstrates that the length scale
of the external space induced by the mass can be identified
by the cutoff.
The functional evolution equation resulting from the 
infinitesimal change of the mass is the generalization
of the Callan-Symanzik equation without relying on
the multiplicative renormalization scheme. Such an extension,
presented below can be used to study non-renormalizable models or effects
close to the cutoff.

When the procedure in the internal space is performed by the help
of the bare action the resulting flow provides a resummation of
the loop expansion. This scheme is specially advantegous for
models with inhomogeneous saddle points because the blocking
does not distorts the tree level structure.

The renormalization group step can be expressed by means of the bare action
or the effective action. The two schemes become
equivalent in the infrared limit. This can be seen easily by remarking that the
loop contributions are suppressed in the infrared limit, where the
ultraviolet cutoff tends to zero with the number of degrees of
freedom. Since the difference between the bare and the effective action
comes from the loop contributions the gradient expansion produces
the same bare and effective action in the vicinity of the infrared
fixed point.

The different standard methods to obtain the renormalization group are
briefly discussed in Section II. The internal space evolution equation,
our new renormalization group scheme is introduced in Section III.
Section IV contains the application of the general strategy of the internal
space renormalization to obtain the generalization of the Callan-Symanzik
equation. The possibility of resumming the loop expansion by solving
the evolution equation is shown in Section V.
The Section VI is for the summary. The appendices
give details about the Legendre transformation and the gradient expansion.

\section{Renormalization group schemes}
In this Section we briefly recapitulate some methods 
the renormalization group equation can be obtained.

The traditional field theoretical methods for
the renormalization group equation are based on the 
simplification offered by placing the
ultraviolet cutoff far away from the scale of the observables.
Such a separation of the scales removes the non-universal pieces of the
renormalized action and the rather complicated blocking 
step can be simplified by retaining the renormalizable coupling constants
only. The underlying formalism is the renormalized perturbation expansion,
in particular the multiplicative renormalization scheme.
The usual perturbative proof of the renormalizability
asserts that the renormalized field and the Green
functions can be written in terms of the bare quantities as
\be
G_n(p_1,\cdots,p_n;g_R(\mu))_R=Z^{-{n\over2}}
\left(g_R,g_B,{\Lambda\over\mu}\right)
G_n(p_1,\cdots,p_n;g_B,\Lambda)_B
\left(1+O\left({p^2\over\Lambda^2}\right)\right),
\label{mult}
\ee
where $\Lambda$ is the cutoff and the renormalized coupling 
constants are defined by some renormalization conditions imposed
at $p^2=\mu^2$. The evolution equation for the
bare and the renormalized coupling constants result from the
requirements
\be\label{barerg}
{d\over d\Lambda}G_R={d\over d\Lambda}Z^{-{n\over2}}G_B=0,
\ee
\be\label{renrg}
{d\over d\mu}G_B={d\over d\mu}Z^{n\over2}G_R=0.
\ee
Note that the non-renormalizable operators can not be treated
in this fashion because the $O(p/\Lambda)$ contributions
are neglected in \eq{mult}. The renormalization of 
composite operators and the corresponding operator
mixing requires the introduction of additional terms 
in the lagrangian. Another aspect of this shortcoming is 
that these methods are useful for the study of the ultraviolet scaling
laws only. The study of the infrared scaling or models where there 
are several non-trivial scaling regimes \cite{glob} require the more 
powerful functional form, introduced below.

The third conventional procedure is the Callan-Symanzik 
equation which is based on the
change of the bare mass parameter,
\be
{d\over dm^2}G_B={d\over dm^2}Z^{n\over2}G_R=Z^{n\over2}Z_{\phi^2}G^{comp}_R
\ee
where $Z_{\phi^2}$ is the renormalization constant for the composite operator
$\phi^2(x)$ and $G^{comp}$ is the Green function with an additional
insertion of $\phi^2(p=0)$. One can convert the mass
dependence inferred from the Callan-Symanzik equation
into the momentum dependence by means of dimensional analysis 
and the resulting expression is usually called a
renormalization group equation. 

The functional generalizations of the renormalization group method
which are based on the infinitesimally small change of the cutoff
allows up to follow the mixing of non-renormalizable operators,
as well, and to trace the evolution close to the cutoff. Another
advantage of these methods is that the renormalization group
equation is either exact or holds in every order of the
loop expansion.

The traditional blocking \cite{wilsrg} in momentum space
yields the Wegner-Houghton equation \cite{wh} which is based on sharp
cutoff in the momentum space. Consider the action with cutoff
$k$ in the derivative expansion,
\be
S_k=\sum\limits_{n=0}^\infty\int d^dxU_k^{(n)}[\phi(x),\partial_\mu]
\ee
where $U_k^{(n)}[\phi(x),\partial_\mu]$ is an n-th order homogeneous polynom
in the gradient operator $\partial_\mu$ and a local functional 
of the field variable. We separate the quadratic part of the 
action in the field, 
\be
S^{quadr}=\hf\int 
d^dxd^dy(\phi(x)-\phi_0(x))G^{-1}(x,y;\phi_0)(\phi(y)-\phi_0(x))
\ee
where
\be
G^{-1}(x,y;\phi_0)={\delta^2S[\phi]\over
\delta\phi(x)\delta\phi(y)}_{|\phi=\phi_0}
\ee
to organize the loop expansion. The blocking step consists of the
elimination of the Fourier components of the field $\phi(x)$
within the shell $k-\dk<p<k$. Let us write
\be
\phi(x)=\tilde\phi(x)+\phi'(x)
\ee
where the first and the second term in non-vanishing in the
Fourier space for $p<k-\dk$ and $k-\dk<p<k$, respectively.
The blocked action can be written as
\bea
e^{-S_{k-\dk}[\tilde\phi]}
&=&\int D[\phi']e^{-S_k[\tilde\phi+\phi']}\nonu
&=&e^{-S_k[\tilde\phi+\psi]-\hf\tr G_k^{-1}[\tilde\phi]+\cdots}
\label{loop}
\eea
where $\psi$ is the saddle point, the trace is within the
functional space spanned by the plane waves $k-\dk<p<k$, 
and the dots stand for 
the higher loop contributions. The density of the modes to 
be eliminated, $\Omega_d\dk k^{d-1}$, $\Omega_d$ being the
solid angle, is chosen to be small. This makes the volume of the
loop integral appearing in the computation of the trace small, too. 
One expects that the n-th loop contributions which contain an n-fold 
integration in this small volume will be suppressed as $\dk\to0$
since the integrand is non-singular in the integration domain. 
A more careful argument shows that the small parameter is actually the
ratio of the modes to be eliminated and left in the system \cite{wh}, 
\cite{glob}
{\em assuming} that the amplitude of each Fourier mode is
of the same order of magnitude\footnote{It is enough to introduce
an infrared cutoff and consider the elimination of the
modes one-by-one when this latter condition fails
to become convinced that the limit $\dk\to0$
{\em can not} always suppresses the higher loop contributions to the
blocking relation. Such a problem shows up when there is a non-trivial
saddle point to the blocking which enhances certain modes \cite{tree}.}. 
With this condition kept in mind the higher loop contributions
are dropped yielding the evolution equation
\be
S_k[\tilde\phi]-S_{k-\dk}[\tilde\phi]=
-\hf\tr G^{-1}[\tilde\phi]\left(1+O\left({\dk\over k}\right)\right)\label{wheq}
\ee
for vanishing saddle point, $\psi=0$.
All we needed in deriving this equation was the availability of the
loop-expansion for the blocking step \eq{loop}. The simplification,
the suppression of the higher loop contribution, results from the
appearance of a new small parameter $\dk/k$ and \eq{wheq} is
valid for weakly coupled models only.

We may call this the scheme the functional form of the bare renormalization
group, \eq{barerg}, since the resulting renormalized trajectory is in the 
space of the cutoff theories. The truncation of the gradient expansion 
\be
S_k=\int d^dx\left[\hf(\partial_\mu\phi)^2(x)+U_k(\phi(x))\right]
\ee
simplifies the functional equation \eq{wheq} into a partial 
differential equation
\be
k\partial_kU_k(\rho)=-{\Omega_d\over2(2\pi)^d}k^d
\ln\left[\left(k^2+\partial_\rho^2U_k(\rho)\right)
\left(k^2+\frac{1}{\rho}\partial_\rho U_k(\rho)\right)^{N-1}\right].\label{senb}
\ee
for an $N$-component field of modulus $\rho$.

The inclusion of the $O(\partial^4)$ higher order terms
in the action is problematic because the sharp momentum 
space cutoff generates non-local interactions \cite{devgrad}.

A similar method was developed for the effective action 
\cite{polc}-\cite{morr} where one introduces the cutoff 
parameter $k$ in the quadratic part of the action 
$G^{-1}(p)\longrightarrow G^{-1}_k(p)$, with
\be
G^{-1}_k(p)=f\left({p\over k}\right)G^{-1}(p),
\label{smoothc}
\ee
where $f(\kappa)$ approaches $1$ and $\infty$
for $\kappa=\infty$ and $\kappa=0$, respectively. We have hope
to generate local, gradient expandable interactions only
if $f(\kappa)$ is an analytic function. Thus locality almost
excludes the complete suppression of the modes, the choice
$f(\kappa)=\infty$ for $\kappa>\kappa_0<\infty$. The exception 
to this rule consists of the periodic, analytic functions, 
\be
G^{-1}_k(p+K)=G^{-1}_k(p)
\ee
where the components of the vector $K$ are integer times $k$.
An example is the lattice regularization where the complete
suppression of the modes outside the first Brillouin-zone
can be achieved without generating non-local interactions.
When the modes
are suppressed gradually one characterizes the suppression
functions by the condition that it changes the most
within the interval $1-\epsilon<\kappa<1+\epsilon$. 
The limit $\epsilon\to0$ is called the sharp cutoff.
The path integration consists of the elimination of the modes
$p>k$ and the evolution equation
\be
\partial_k\int D[\phi]e^{-S_k[\phi]+\int_xj_x\phi_x}
=-\int D[\phi]\partial_kS_k[\phi]
e^{-S_k[\phi]+\int_xj_x\phi_x}\label{exww}
\ee
is obtained for the generator functional.
This relation is finally converted into an equation for 
the effective action. Since the effective action
for the modes $p>k$ is followed in this manner, these
methods appear the analogues of \eq{renrg}, the evolution
equation for the running coupling constants.

The application of \eq{exww} is a safe non-perturbative step only for 
a differentiable suppression function $f(\kappa)$. When the $f(\kappa)=\infty$
for finite values of $\kappa$ or when $f(\kappa)$ has
infinite derivative (say $\epsilon\to0$) then \eq{exww} is invalid. Instead
one has to consider the finite difference equation $k\to k-\dk$
and the corresponding functional Taylor expansion brings 
back the assumption concerning the applicability of
the loop-expansion. Thus \eq{exww} produces an exact
relation for smoothly suppressed modes only. 

The internal space renormalization group method presented
in the next section generalises the third traditional way of
introducing the renormalization group, the Callan-Symanzyk
equation by controlling the amplitude of the fluctuations
in the internal space only. The coresponding suppression function 
can be choose as smooth as possible, $f(\kappa)=\lambda$, in order
to suppress the higher order derivative terms generated by the
blocking.

\section{Evolution equation}
Our goal is to obtain the effective action $\Gamma[\phi]$
of the Euclidean model defined by the action $S_B[\phi]$, by reducing
the renormalization group strategy into an algorithm to
solve the theory. The connection to the external scale dependence 
will be considered later. The usual Legendre transformation 
(c.f. Appendix A.) yields
\be
e^{W[j]}=\int D[\phi]e^{-S_B[\phi]+\int_xj_x\phi_x}
\ee
and
\be
W[j]+\Gamma[\phi]=\int_xj_x\phi_x=j\cdot\phi,
\ee
the source $j$ is supposed to be expressed in terms of 
\be
\phi_x={\delta W[j]\over\delta j_x}.
\label{phij}
\ee
A cutoff $\Lambda$ is assumed implicitly in the path integral and 
$S_B[\phi]$ stands for the bare, cutoff action. 

A sharp cutoff could be introduced in the internal space as
\be
Z_\Phi=\prod\limits_x\int\limits_{-\Phi}^\Phi d\phi_xe^{-S[\phi]},
\ee
but it leads to unwanted complications and 
will be replaced by a smooth cutoff as follows. We modify the bare action 
\be
S_B[\phi]\longrightarrow S_\la[\phi]=\la S_s[\phi]+S_B[\phi]
\ee
in such a manner that the model with $\la\to\infty$ be soluble.
This is achieved by requiring that the minimum $\phi_\infty$
of the local functional $S_s[\phi]\ge0$, 
is nondegenerate. In fact, the path integral for $\lambda=\infty$ 
contains no fluctuations. The role of the new piece in the action
is to suppress the fluctuations around $\phi_\infty$
for large value of $\lambda$ and render the model perturbative. 
The control of the amplitude of the fluctuations offered by the
parameter $\lambda$ is considered as a smooth cutoff
in the internal space. The difference between our strategy
and the other methods is that the internal space cutoff
appears multiplicatively in the action and influences
all modes simultaneously while the external space cutoff is modifying the
momentum dependence. In other words, it
is the order the different fluctuations are treated as the cutoff
is decreased what distinguishes the internal and the
external space renormalization group schemes.

We plan to follow the $\la$ dependence of the effective action
by integrating out the functional differential equation
\be
\partial_\la\Gamma={\cal F}_\la[\Gamma]\label{evol}
\ee
from the initial condition 
\be
\Gamma_{\la_{init}}[\phi]=\la_{init}S_s[\phi]+S_B[\phi],
\ee
imposed at $\la_{init}\approx\infty$ to $\la=0$. \eq{evol} can be
interpreted as a generalization of the
Callan-Symantzik equation becaue both generate
a one-parameter family of different theories organized according
to the strength of the quantum fluctuations
\footnote{Note that the inverse mass is proportional to the 
amplitude of the fluctuations.}. So long 
as the parameter $\la$ introduces a renormalization scale, $\mu(\la)$,
the trajectory $\Gamma_{\la(\mu)}[\phi]$ in the effective action
space can be thought as a renormalized trajectory. Another
way to interpret \eq{evol} is to consider its integration as a
method which builds up the fluctuations of the model with 
$\la=0$ by summing up the effects of increasing the fluctuation 
strength infinitesimally, $\la\to\la-\Delta\la$. Notice that the
gauge invariance of the evolution equation \eq{evol} is obvious
when $S_s$ is gaue invariant. The gradient expansion
is compatible with \eq{evol} if the suppression is sufficiently
smooth in the momentum spcae, i.e. $S_s$ is a local functional.

The starting point to find ${\cal F}_\la[\Gamma]$ is
the relation
\be
\partial_\la\Gamma[\phi]=-\partial_\la W[j]
-\frac{\delta W[j]}{\delta j}.\partial_\lambda j
+\partial_\lambda j.\phi=-\partial_\la W[j],
\ee
$\lambda$ and $\phi$ being the independent variables. This relation
will be used together with
\be
\partial_\la W[j]=-e^{-W[j]}\int D[\phi]S_s[\phi]
e^{-\la S_s[\phi]-S_B[\phi]}=-e^{-W[j]}
S_s\left[{\delta\over\delta j}\right]e^{W[j]}.
\ee
It is useful to perform the replacement
\be
\Gamma[\phi]\longrightarrow\la S_s[\phi]+\Gamma[\phi]
\ee
which results the evolution equation
\be
\partial_\la\Gamma[\phi]=e^{-W[j]}
S_s\left[{\delta\over\delta j}\right]e^{W[j]}-S_s[\phi].
\label{evolk}
\ee

The next question is the choice of $S_s[\phi]$. 
The simplest is to use a quadratic suppression term,
\be
S_s[\phi]=\hf\int_{x,y}\phi_x\cm_{x,y}\phi_y
=\hf\phi\cdot\cm\cdot\phi.
\label{qsup}
\ee
If a gauge symmetry should be preserved then such an $S_s[\phi]$
is not acceptable and 
\be
S_s[\phi]=S_B[\phi]
\label{hsup}
\ee
is the most natural choice.
The evolution equation \eq{evolk} then sums up the loop expansion and
produces the dependence in $\hbar$. We return now to the case of a simple
scalar field without local symmetry, \eq{qsup}. The 
corresponding evolution equation can be ontained from \eq{evolk},
and considering the relation \eq{propi} between the functional derivatives of
$W[j]$ and $\Gamma[\phi]$,
\bea\label{evolkk}
\partial_\la\Gamma[\phi]&=&\hf\int_{x,y}\cm_{x,y}\left[
W^{(2)}_{x,y}+\phi_x\phi_y\right]
-\hf\int_{x,y}\phi_x\cm_{x,y}\phi_y\nonu
&=&\hf\int_{x,y}\cm_{x,y}\left[\Gamma^{(2)}_{x,y}+\la\cm_{x,y}
\right]^{-1}
\eea
where the functional derivatives are denoted by
\be
\Gamma^{(n)}_{x_1,\cdots,x_n}=
{\delta^n\Gamma[\phi]\over\delta\phi_{x_1}\cdots\delta\phi_{x_n}}.                     
\ee                                                                                    
\eq{evolkk} reads in an operator notation
\be
\partial_\la\Gamma[\phi]=\hf\mbox{Tr}\left\{\cm\cdot\left[
\la\cm+\Gamma^{(2)}\right]^{-1}\right\}, 
\ee
We should bear in mind that $\Gamma^{(n)}_{x_1,\cdots,x_n}$ 
remains a functional of the field $\phi_x$. 

It is illuminating to compare
this result with the evolution equations presented in refs.
\cite{polc}-\cite{tetra} what one can obtain by means of 
\eq{smoothc} and \eq{evolk},
\be
\partial_k\Gamma[\phi]=\hf\mbox{Tr}\left\{\partial_kG^{-1}_k\cdot\left[
G^{-1}_k+\Gamma^{(2)}\right]^{-1}\right\}.
\label{pwm}
\ee
The role of $S_s$ is played here by the propagator $G_k(p)$ which 
contains the external space scale parameter $k$ to control
the supression of the fluctuations. 
The formal similarity with \eq{evolkk}
reflects that the different schemes agree
in "turning on" the fluctuations in infinitesimal steps. 
But the internal space scheme operates with a suppression term which is
regular in or even independent of the momentum.

The evolution equation can be converted into
a more treatable form by the means of the gradient
expansion,
\be\label{gradexp}
\Gamma[\phi]=\int_x\left\{\frac{1}{2}Z_x
(\partial_\mu\phi_x)^2+U_x+O(\partial^4)\right\}
\ee
where the notation $f_x=f(\phi_x)$ was introduced. This ansatz gives
\bea
\Gamma^{(1)}_{x_1}&=&-\hf Z^{(1)}{x_1}(\partial_\mu\phi_{x_1})^2
-Z_{x_1}\Box\phi_{x_1}+U^{(1)}_{x_1}\\
\Gamma^{(2)}_{x_1,x_2}
&=&-\hf\delta_{x_1,x_2}Z^{(2)}_{x_1}(\partial_\mu\phi_{x_1})^2
-\partial_\mu\delta_{x_1,x_2}Z^{(1)}_{x_1}\partial_\mu\phi_{x_1}\nonu
&&-\delta_{x_1,x_2}Z^{(1)}_{x_1}\Box\phi_{x_1}
-\Box\delta_{x_1,x_2}Z_{x_1}+U^{(2)}_{x_1}\nonumber
\eea
where the $f^{(n)}(\phi)=\partial^n_\phi f(\phi)$.
Such an expansion is unsuitable for $W[j]$ due to the strong
non-locality of the propagator but might be more
successful for the effective action where the one-particle
irreducible structure and the removal of the propagator at the 
external legs of the contributing diagrams strongly reduce
the non-local effects. The replacement of this ansatz into 
\eq{evolk} gives (c.f. Appendix B.)
\bea\label{evoluz}
\partial_\la U_\la(\phi)&=&\hf\int_p
{{\cal M}(p)\over\la {\cal M}(p)+Z_\la(\phi)p^2+U_\la^{(2)}(\phi)}\nonu
\partial_\la Z_\la(\phi)&=&\hf\int_p
{\cal M}(p)\left[-\frac{Z_\la^{(2)}(\phi)}
{\left(\la {\cal M}(p)+Z_\la(\phi)p^2+U_\la^{(2)}(\phi)\right)^2}\right.\nonu
&+&2Z_\la^{(1)}(\phi)\frac{2\left(Z_\la^{(1)}(\phi)p^2+U_\la^{(3)}(\phi)\right)
+Z_\la^{(1)}(\phi)p^2/d}
{\left(\la {\cal M}(p)+Z_\la(\phi)p^2+U_\la^{(2)}(\phi)\right)^3}\nonu
&-&\frac{\left(Z_\la^{(1)}(\phi)p^2+U_\la^{(3)}(\phi)\right)^2
\left(\la\Box {\cal M}(p)+2Z_\la(\phi)\right)}
{\left(\la {\cal M}(p)+Z_\la(\phi)p^2+U_\la^{(2)}(\phi)\right)^4}\nonu
&-&\frac{4}{d}Z_\la^{(1)}(\phi)
\left(Z_\la^{(1)}(\phi)p^2+U_\la^{(3)}(\phi)\right)
\frac{\left(\la p_\mu \partial_\mu {\cal M}(p)+2Z_\la(\phi)p^2\right)}
{\left(\la {\cal M}(p)+Z_\la(\phi)p^2+U_\la^{(2)}(\phi)\right)^4}\nonu
&+&\left.\frac{2}{d}\frac{\left(Z_\la^{(1)}(\phi)p^2+U_\la^{(3)}(\phi)\right)^2
\left(\la\partial_\mu {\cal M}(p)+2Z_\la(\phi)p_\mu\right)^2}
{\left(\la {\cal M}(p)+Z_\la(\phi)p^2+U_\la^{(2)}(\phi)\right)^5}\right]
\eea
where $\int_p=\int{d^dp\over(2\pi)^d}$ and we assumed that $\partial_\mu\cm(p)$
is proportional to $p_\mu$. Since the order the fluctuations are
treated is different for the internal and the external space methods,
\eq{evoluz} is different, as well, from the evolution equations
of \cite{wh}-\cite{tetra}. The final solution, correpsonding to $\lambda=0$
in our case and $k=0$ in the others, is the effective potential expressed in
terms of the bare and the renormalized coupling constants, respectively.

\section{Mass dependence}
The simplest choice is $\la=m^2$ with
\be
\cm_{x,y}=\delta_{x,y}
\ee
which minimizes strength of the higher order derivative terms generated
during the evolution by being a momentum independent suppression mechanism.
The evolution equation
is the functional differential renormalization group version 
of the Callan-Symanzik equation,
\be
\partial_{m^2}\Gamma[\phi]=\hf\mbox{Tr}\left[m^2\delta_{x,y}
+\Gamma^{(2)}_{x,y}\right]^{-1}.
\ee
The projection of this functional equation onto the gradient expansion 
ansatz gives
\bea\label{csrg}
\partial_{m^2}U(\phi)&=&\hf\int_p
{1\over Z(\phi)p^2+m^2+U^{(2)}(\phi)}\nonu
\partial_{m^2}Z(\phi)&=&\hf\int_p
\left[-\frac{Z^{(2)}(\phi)}
{\left(Z(\phi)p^2+m^2+U^{(2)}(\phi)\right)^2}\right.\nonu
&~&~~~~~~~+2Z^{(1)}(\phi)\frac{p^2/dZ^{(1)}(\phi)+
2\left(Z^{(1)}(\phi)p^2+U^{(3)}(\phi)\right)}
{\left(Z(\phi)p^2+m^2+U^{(2)}(\phi)\right)^3}\nonu
&~&~~~~~~~-2Z(\phi)\frac{\left(Z^{(1)}(\phi)p^2+U^{(3)}(\phi)\right)^2}
{\left(Z(\phi)p^2+m^2+U^{(2)}(\phi)\right)^4}\nonu
&~&~~~~~~~-\frac{8p^2}{d}Z(\phi)Z^{(1)}(\phi)
\frac{\left(Z^{(1)}(\phi)p^2+U^{(3)}(\phi)\right)}
{\left(Z(\phi)p^2+m^2+U^{(2)}(\phi)\right)^4}\nonu
&~&~~~~~~~+\left.\frac{8p^2}{d}Z^2(\phi)\frac{\left(Z^{(1)}(\phi)p^2+U^{(3)}
(\phi)\right)^2}
{\left(Z(\phi)p^2+m^2+U^{(2)}(\phi)\right)^5}\right]
\eea
It is important to bear in mind that we are dealing here
with a well regulated theory and that the procedure described here 
does not aim at removing the external space cutoff $\Lambda$ which remains an
important external parameter. In the usual, external space
renormalization schemes the blocking provides the cutoff
and the regulator for the models. In our case the cutoff
in the internal space can not replace the usual external space
cutoff $\Lambda$ and this latter must properly be implemented from the 
very beginning, c.f. the remark after \eq{phij}. In this case
the loop integrals occuring in the evolution equations
are finite.

Let us now simplify the differential equation for $U(\phi)$ and 
$Z(\phi)$ by integrating over $p$ in \eq{csrg} with sharp
momentum cutoff $\Lambda$\footnote{
The sharp momentum space cutoff generates nonlocal interactions.
Since these nonlocal contributions come from the surface terms of the loop
integrals they are suppressed in a renormalizable theory when $\Lambda$
is kept large. Thus the gradient expansion anzatz can be justified
for the evolution \eq{csrg}.}
in four dimensions,
\bea\label{diffuz}
\partial_{m^2}U(\phi)&=&\frac{1}{32\pi^2Z(\phi)}
\left[\Lambda^2-{m^2+U^{(2)}(\phi)\over Z(\phi)}
\ln\left(1+\frac{Z(\phi)\Lambda^2}{m^2+U^{(2)}
(\phi)}\right)\right]\nonu
\partial_{m^2}Z(\phi)&=&\frac{1}{32\pi^2Z(\phi)}\left[\frac{1}{Z^2(\phi)}
\left(\frac{5}{2}\left(Z^{(1)}(\phi)\right)^2-Z(\phi)Z^{(2)}(\phi)\right)
\ln\left(1+\frac{Z(\phi)\Lambda^2}{m^2+U^{(2)}(\phi)}\right)\right.\nonu
&~&~~~~~~~~~~~~~~+\frac{1}{Z^2(\phi)}\left(Z(\phi)Z^{(2)}(\phi)-\frac{43}{12}
\left(Z^{(1)}(\phi)\right)^2\right)\nonu
&~&~~~~~~~~~~~~~~+\left.\frac{1}{Z(\phi)}\frac{Z^{(1)}(\phi)
U^{(3)}(\phi)}{\left(m^2+U^{(2)}(\phi)\right)}
-\frac{1}{6}\frac{\left(U^{(3)}(\phi)\right)^2}
{\left(m^2+U^{(2)}(\phi)\right)^2}\right]
\eea
In the approximation $Z=1$ we obtain
\be
\partial_{m^2}U(\phi)=-{m^2+U^{(2)}(\phi)\over32\pi^2}
\ln\left(1+{\Lambda^2\over m^2+U^{(2)}(\phi)}\right)
\ee
after removing a field independent term. In order to simplify 
the scaling relations we consider the regime $m^2\gg U^{(2)}$, where
\be\label{intas}
\partial_{m^2}U(\phi)=-{1\over32\pi^2}
\ln\left({m^2+\Lambda^2\over m^2}\right)U^{(2)}(\phi)
\ee
The asymptotic scaling formula of the
bare renormalization group \eq{senb} for $d=4$ in the same regime is
\be\label{whas}
k\partial_kU(\phi)=-{k^2\over16\pi^2}U^{(2)}(\phi).
\ee
The evolutions \eq{intas} and \eq{whas} agree up to an overall constant 
in the scale parameter if
\be
\frac{dk^2}{dm^2}=\ln\left({m^2+\Lambda^2\over m^2}\right)
\label{match}
\ee
can be considered as constant, i.e. $m^2$ and $k$ independent. 
We obtain in this manner the usual justification of calling
the Callan-Symanzik equation a renormalization group method,
the equivalence of the scales in the internal and the external 
spaces in the ultraviolet scaling regime $m^2\gg U^{(2)}$ up to
non-universal contributions. The equivalence of the scales and the
elimination of the non-universal contributions requires that the cutoff 
should be far above the mass, $m^2\ll\Lambda^2$.

The non-vanishing anomalous dimension sets up the relation between the
internal and external space scaling. In fact, when $Z\not=1$ 
the relation \eq{match} becomes field dependent according to the
first equation of \eq{diffuz}. It is worthwhile comparing what
\eq{diffuz} gives in the asymptotical regime $m^2\gg U^{(2)}$,
\be
\partial_{m^2}Z_{m^2}(\phi)=-\frac{1}{32\pi^2Z_{m^2}^3(\phi)}
\ln\left(\frac{Z_{m^2}(\phi)\Lambda^2+m^2}{m^2}\right)
\left[Z_{m^2}(\phi)Z_{m^2}^{(2)}(\phi)-\frac{5}{2}
\left(Z_{m^2}^{(1)}(\phi)\right)^2\right]
\label{inza}
\ee
with the prediction of the Wegner-Houghton equation.
A possible attempt to save the gradient expansion with sharp cutoff
for the latter is the following: The contributions to the coefficient
functions of the gradient, such as $Z_k(\phi)$, come from
taking the derivative of the loop integral, the trace in the
second equation of \eq{loop}, with respect to the momentum of the
infrared background field $\tilde\phi(x)$. 
There are two kind of contributions,
one which comes form the derivative of the integrand, another
from the external momentum dependence of the limit of the 
integration. It is easy to verify that the $\epsilon$-dependent 
non-local contributions come form the second types only \cite{morr}.
Thus one may consider the approximation where these contributions
are simply neglected, assuming a cancellation mechanism
between the successive blocking steps. The result is, for 
$k^2\gg U^{(2)}_k(\phi)$, c.f. Appendix C,
\be\label{equazk}
k\partial_kZ_k(\phi)=-\frac{k^2}{32\pi^2Z_k^2(\phi)}
\left[2Z_k(\phi)Z_k^{(2)}(\phi)-\frac{5}{2}
\left(Z_k^{(1)}(\phi)\right)^2\right].
\ee
The formal similarity between the two different schemes, \eq{inza} and
\eq{equazk}, can be considered as a measure of the cancellation of 
the non-local terms evoked above.

The beta-functions of the coupling constants $g_n$
and $z_m$ introduced as
\be
U(\phi)=\sum_n{g_n\over n!}\phi^n,~~~
Z(\phi)=\sum_n{z_n\over n!}\phi^n
\ee
are of the form
\bea
\beta_n&=&m^2\partial_{m^2}g_n\nonu
&=&C_dm^2\frac{\partial^n}{\partial\phi^n}\int_0^{\Lambda^2} dyy^{{d\over2}-1}
{1\over Z(\phi)y+m^2+U^{(2)}(\phi)}\nonu
\gamma_n&=&m^2\partial_{m^2}z_n\nonu
&=&
C_dm^2\frac{\partial^n}{\partial\phi^n}\int_0^{\Lambda^2} dyy^{{d\over2}-1}
\left[-\frac{Z^{(2)}(\phi)}
{\left(Z(\phi)y+m^2+U^{(2)}(\phi)\right)^2}\right.\nonu
&+&2Z^{(1)}(\phi)\frac{y/dZ^{(1)}(\phi)+
2\left(Z^{(1)}(\phi)y+U^{(3)}(\phi)\right)}
{\left(Z(\phi)y+m^2+U^{(2)}(\phi)\right)^3}
-2Z(\phi)
\frac{\left(Z^{(1)}(\phi)y+U^{(3)}(\phi)\right)^2}
{\left(Z(\phi)y+m^2+U^{(2)}(\phi)\right)^4}\nonu
&-&\left.\frac{8y}{d}Z(\phi)Z^{(1)}(\phi)
\frac{\left(Z^{(1)}(\phi)y+U^{(3)}(\phi)\right)}
{\left(Z(\phi)y+m^2+U^{(2)}(\phi)\right)^4}
+\frac{8y}{d}Z^2(\phi)
\frac{\left(Z^{(1)}(\phi)y+U^{(3)}
(\phi)\right)^2}
{\left(Z(\phi)y+m^2+U^{(2)}(\phi)\right)^5}\right]
\eea
with $C_d=\Omega_d/2(2\pi)^d$. The integration over $y$ produces
simple expressions for $\beta_n$ and $\gamma_n$ in terms of 
the coupling constants $g_m$ and $z_m$. The simultaneous integration 
of this set of equations produces the solution of the evolution 
equation \eq{csrg}.

It is instructive to consider the solution in the 
independent mode approximation where the $m^2$ dependence 
is ignored in the integrals, $U(\phi)=U_B(\phi)$ and $Z(\phi)=1$.
We get
\bea
U_{eff}(\phi)&=&U_B(\phi)
+\hf\int^0_{M^2}dm^2\int_p
{1\over p^2+m^2+U_B^{(2)}(\phi)}\nonu
&=&U_B(\phi)+\hf\int_p\ln
[p^2+U_B^{(2)}(\phi)]+O(M^{-2}),\nonumber
\eea
which reproduces the usual one-loop effective potential for
$M\gg\Lambda$. For the kinetic term, the integration of (\ref{diffuz})
in the same approximation leads to
\bea
Z_{eff}(\phi)&=&1-\frac{1}{192\pi^2}\int_{M^2}^0dm^2
\frac{\left(U_B^{(3)}(\phi)\right)^2}{\left(m^2+U_B^{(2)}(\phi)\right)^2}\nonu
&=&1+\frac{1}{192\pi^2}\frac{\left(U_B^{(3)}(\phi)\right)^2}
{U_B^{(2)}(\phi)}+O(M^{-2}),
\eea
for $d=4$ which reproduces the one-loop solution found in \cite{fraser}.
The agreement between the independent mode approximation to the
internal-space renormalization group equation and the one-loop solution
is expected not only because the right hand side of \eq{evolkk} is $\ord(\hbar)$
but because the one-loop contribution to the gamma function is universal,
scheme independent. But this agreement does not hold beyond $\ord(\hbar)$
as indicated by the imcompatibility of \eq{inza} and \eq{equazk}.

\section{$\hbar$ dependence}
It may happen that the quadratic suppression is not well suited
to a problem. In the case $S_B[\phi]$ possesses local symmetries which should be
preserved then the choice \eq{hsup} is more appropriate.
The application of our procedure for a gauge model can for
example be based on the choice
\bea
S_B[A]&=&-{1\over4g^2_B}\int dxF_{\mu\nu}^aF^{\mu\nu a}+S_{gf}[A],\nonu
S_s[A]&=&-{1\over4g^2_B}\int dxF_{\mu\nu}^aF^{\mu\nu a},
\eea
where $S_{gf}$ contains the gauge fixing terms and on the application of 
a gauge invariant regularization scheme. As mentioned after eq.
\eq{csrg} we need a regulator to start with in order to follow the dependence
on the amplitude of the fluctuations. One may use lattice, analytic
(asymptotically free models) or Pauli-Villars (QED) regulator to
render \eq{evolk} well defined.
The explicit gauge invariance of $S_s[A]$ which was
achieved by suppressing the gauge covariant field strength instead
of the gauge field itself makes obvious the independence of
the resulting flow for the gauge invariant part of the action
from the choice of the gauge, $S_{gf}$.
When a non-trivial saddle point appears in the blocking
step then it may develop a discontinuous evolution. The choice
\eq{hsup} makes the saddle point approximatively
``renormalization group invariant''. 

We present the evolution equation for the $\phi^4$ model with quartic
supression,
\be
S_B[\phi]=S_s[\phi]=\int_x\left[\hf(\partial_\mu\phi_x)^2
+{g_2\over2}\phi_x^2+{g_3\over3!}\phi_x^3+{g_4\over4!}\phi_x^4\right].
\ee
The similarity of this scheme with the loop expansion suggests the
replacement 
\be
{1\over\hbar}=1+\la=1+{1\over g},
\ee
which yields the evolution equation
\be
\partial_g\Gamma[\phi]=-{1\over g^2}e^{-W[j]}
S_s\left[{\delta\over\delta j}\right]e^{W[j]}+{1\over g^2}S_s[\phi].
\ee
The integration of the evolution equation
from $g_{in}=0$ to $g_{fin}=\infty$ coresponds
to the resummation of the loop expansion, i.e. the integration
between $\hbar_{in}=0$ and $\hbar_{fin}=1$.

The gradient expansion anstatz \eq{gradexp} with $Z=1$ gives
\bea\label{phin}
\partial_gU(\phi)&=&
-{1\over g^2}\Biggl\{\hf\int_p (p^2+g_2)G(p)\\
&&+{g_3\over3!}\biggl[3\phi\int_pG(p)
-\int_{p_1,p_2}G(p_1)G(p_2)G(-p_1-p_2)
\left(U^{(3)}(\phi)+g^{-1}(g_3+g_4\phi)\right)\biggr]\nonu
&&+{g_4\over4!}\biggl[3\left(\int_pG(p)\right)^2+6\phi^2\int_pG(p)\nonu
&&~~~~~~~-\int_{p_1,p_2,p_3}G(p_1)G(p_2)G(p_3)G(-p_1-p_2-p_3)
\left(U^{(4)}(\phi)+g^{-1}g_4\right)\nonu
&&~~~~~~~-3\int_{p_1,p_2,p_3}G(p_1)G(p_2)G(p_3)G(-p_1-p_2)G(-p_1-p_2-p_3)\nonu
&&~~~~~~~~~~~~~~~~~~~~~~~\times\left(U^{(3)}(\phi)
+g^{-1}(g_3+g_4\phi)\right)^2\nonu
&&~~~~~~~-4\phi\int_{p_1,p_2}G(p_1)G(p_2)G(-p_1-p_2)
\left(U^{(3)}(\phi)+g^{-1}(g_3+g_4\phi)\right)\biggr]\Biggr\},
\nonumber
\eea
where we used the fact that the Fourier transform of the 1PI amplitude 
for $n\ge3$ and $Z=1$ is
\be
\int_{x_1,\cdots,x_n}e^{i(p_1\cdot x_1+\cdots+p_n\cdot x_n)}
\Gamma^{(n)}(x_1,\cdots,x_n)=
(2\pi)^d\delta(p_1+\cdots+p_n)U^{(n)}(\phi).
\ee
The propagator in the presence of the homogeneous background field $\phi$
is given by
\be
G(p)=\left[p^2+U^{(2)}(\phi)+g^{-1}\left(p^2+g_2+g_3\phi
+{g_4\over 2}\phi^2\right)\right]^{-1}.
\ee
Since the momentum dependence in the right hand side of \eq{phin} 
is explicit and simple the one, two and three loop integrals can be carried out
easily by means of the standard methods. The successive derivatives of the
resulting expression with respect to $\phi$ yield the renormalization group
coefficient functions.

The use of internal space renormalization discribed in this section 
shows that this method can be generalized to any kind of action $S_s$
and not only to a quadratic suppression term, as shown in the previous
sections.

\section{Summary}
The strategy of the renormalization group is developed further in this paper. 
Instead of following the evolution of the coupling constants corresponding to
the same physics our renormalization group flow sweeps through models
with different dynamics. The parameter of the flow is the scale of the
quantum or thermal fluctuations. The result is an exact functional 
differential equation for the effective action. As a special case the
functional generalization of the Callan-Symanzik equation is recovered.
A different choice of the "blocking transformation" allows us to control $\hbar$ 
and the resulting flow amounts to the resummation of the loop expansion.

Our scheme can be considered as a renormalization group method in the 
internal space. The similarity of the renormalization group flow
in the external and the internal space is shown for the local potential,
the zero momentum piece of the effective action. The difference
between the two schemes is the source of the anomalous dimension
and it appears in the momentum dependent parts of the effective action.

The novel feature of the method is its manifest gauge 
invariance. This is achieved by the possibility of characterizing the modes in the internal 
space while the gauge transformations are carried out in the external space.

The integration of the internal space evolution equation
provides an algorithm to solve models in a manner similar
to the traditional renormalization group method. 
The only truncation is done in the gradient expansion of the effective 
action, in the Taylor expansion of the 1PI functions in the momentum.
We believe that this procedure is an interesting alternative to the stochastic
solution method of the lattice regulated models. The drawback is that
it is rather cumbersome, thought possible in principle, to increase the
precision in the momentum and going to higher orders in the gradient
expansion. The advantage is that it can be cast in infinite, continuous 
space-time equipped with Minkowski metric.\\

{\em Note added in proof:} After this work has been completed we learned
that a method presented for gauge models in ref. \cite{simi} is similar to ours
in the case of mass dependence (section IV). \cite{simi} gives 
a loop expanded solution of the exact equation, whereas our solution 
is built in the framework of the derivative expansion. Finally,
our approach can be generalized to any kind of suppression action $S_s$
which is compatible with the symetries as shown in section V.

\begin{appendix}
\section{Legendre transformation}
We collect in this Appendix the relations between 
the derivatives of the generator functional $W[j]$ and
$\Gamma[\phi]$ used in obtaining the evolution equations
for $\Gamma$.

We start with the definitions
\be
W[j]+\Gamma[\phi]+\la S_s[\phi]=j\cdot\phi,
\ee
and
\be
\phi_x=W^{(1)}_x.\label{phj}
\ee
The first derivative of $\Gamma$ gives the
inversion of \eq{phj},
\be
\Gamma^{(1)}_x=j_x-\lambda S_{s,x}^{(1)}.
\label{jph}
\ee
The second derivative is related to the propagator
$W^{(2)}_{x_1,x_2}=G_{x_1,x_2}$ 
\be
\Gamma^{(2)}_{x_1,x_2}={\delta j_{x_1}\over\delta\phi_{x_2}}
-\la S_{s,x_1,x_2}^{(2)}=G_{x_1,x_2}^{-1}-\la S_{s,x_1,x_2}^{(2)}.
\label{propi}
\ee
The third derivative is obtained by differentiating \eq{propi},
\be
\Gamma^{(3)}_{x_1,x_2,x_3}=
-\int_{y_1,y_2,y_3}G^{-1}_{x_1,y_1}G^{-1}_{x_2,y_2}G^{-1}_{x_3,y_3}
W^{(3)}_{y_1,y_2,y_3}-\la S_{s,x_1,x_2,x_3}^{(3)}.
\ee
The inverted form of this equation is
\be
W^{(3)}_{x_1,x_2,x_3}=-\int_{y_1,y_2,y_3}G_{x_1,y_1}G_{x_2,y_2}
G_{x_3,y_3}\left(\Gamma^{(3)}_{y_1,y_2,y_3}+\la S_{s,y_1,y_2,y_3}^{(3)}\right).
\ee
The further derivation gives
\bea
\Gamma^{(4)}_{x_1,x_2,x_3,x_4}&=&
\int_{y_1,y_2,y_3,y_4,z_1,z_2}\biggl[
G^{-1}_{x_1,y_1}G^{-1}_{x_2,y_2}G^{-1}_{x_3,y_3}G^{-1}_{x_4,y_4}
W^{(4)}_{y_1,y_2,y_3,y_4}\nonu
&&+G^{-1}_{x_1,y_1}G^{-1}_{x_2,y_2}W^{(3)}_{y_1,y_2,z_1}
G^{-1}_{z_1,z_2}W^{(3)}_{z_2,y_3,y_4}
G^{-1}_{x_3,y_3}G^{-1}_{x_4,y_4}\nonu
&&+G^{-1}_{x_3,y_3}G^{-1}_{x_2,y_2}W^{(3)}_{y_3,y_2,z_1}
G^{-1}_{z_1,z_2}W^{(3)}_{z_2,y_1,y_4}
G^{-1}_{x_1,y_1}G^{-1}_{x_4,y_4}\nonu
&&+G^{-1}_{x_1,y_1}G^{-1}_{x_4,y_4}W^{(3)}_{y_1,y_4,z_1}
G^{-1}_{z_1,z_2}W^{(3)}_{z_2,y_3,y_2}
G^{-1}_{x_3,y_3}G^{-1}_{x_2,y_2}\biggr]\nonu
&&-\la S_{s,x_1,x_2,x_3,x_4}^{(4)}.
\eea
Its inversion expresses the four point connected Green
function in terms of the 1PI amplitudes,
\bea
W^{(4)}_{x_1,x_2,x_3,x_4}
&=&\int_{y_1,y_2,y_3,y_4,z_1,z_2}\biggl[
G_{x_1,y_1}G_{x_2,y_2}G_{x_3,y_3}G_{x_4,y_4}
\left(\Gamma^{(4)}_{y_1,y_2,y_3,y_4}+\la S_{s,y_1,y_2,y_3,y_4}^{(4)}\right)\\
&&-G_{x_1,y_1}G_{x_2,y_2}\left(\Gamma^{(3)}_{y_1,y_2,z_1}+\la 
S_{s,y_1,y_2,z_1}^{(3)}\right)
G_{z_1,z_2}\left(\Gamma^{(3)}_{z_2,y_3,y_4}+\la S_{s,z_2,y_3,y_4}^{(3)}\right)
G_{x_3,y_3}G_{x_4,y_4}\nonu
&&-G_{x_3,y_3}G_{x_2,y_2}\left(\Gamma^{(3)}_{y_3,y_2,z_1}+\la 
S_{s,y_3,y_2,z_1}^{(3)}\right)
G_{z_1,z_2}\left(\Gamma^{(3)}_{z_2,y_1,y_4}+\la S_{s,z_2,y_1,y_4}^{(3)}\right)
G_{x_1,y_1}G_{x_4,y_4}\nonu
&&-G_{x_1,y_1}G_{x_4,y_4}\left(\Gamma^{(3)}_{y_1,y_4,z_1}+\la 
S_{s,y_1,y_4,z_1}^{(3)}\right)
G_{z_1,z_2}\left(\Gamma^{(3)}_{z_2,y_3,y_2}+\la S_{s,z_2,y_3,y_2}^{(3)}\right)
G_{x_3,y_3}G_{x_2,y_2}\biggr].\nonumber
\nonumber
\eea

\section{Evolution equation in the internal space}
We give here some details on the computation of (\ref{evoluz}).
To get the evolution equation of the potential part of the gradient expansion
(\ref{gradexp}), one has to take a homogeneous field $\phi=\phi_0$
in (\ref{evolkk}). But to distinguish the kinetic contribution from the 
potential one, a non homogeneous field $\phi(x)=\phi_0+\eta(x)$ is needed, as
well. Let $k$ be the momentum where the field $\eta$ is non-vanishing. 
Then the effective action can be written as
\be
\Gamma[\phi]=V_dU_\la(\phi_0)+\hf\int_q\tilde\eta(q)\tilde\eta(-q)
\left(Z_\la(\phi_0)q^2+U_\la^{(2)}(\phi_0)\right)+\ord(\tilde\eta^3,k^4)
\ee
where $V_d$ is the spatial volume. Thus we need the second derivative of 
the effective action in (\ref{evolkk}) up to the second order in 
$\tilde\eta$ to identify the different contributions. The terms 
independent of $\tilde\eta$
give the equation for $U_\la$ and the ones proportional to $k^2\tilde\eta^2$
the equation for $Z_\la$. The contributions proportional to $\tilde\eta^2$ but 
independent of $k$ yield an equation for $U_\la^{(2)}$ which must
be consistent with the equation for $U_\la$. The result is 
\bea
\Gamma^{(2)}_{p_1,p_2}&=&\left[Z_\la(\phi_0)p_1^2+U_\la^{(2)}(\phi_0)\right]
\delta(p_1+p_2)\\
&+&\int_q\tilde\eta(q)\left[Z_\la^{(1)}(\phi_0)(p_1^2+q^2+qp_1)+U_\la^{(3)}(\phi
_
0)\right]
\delta(p_1+p_2+q)\nonu
&+&\hf\int_{q_1,q_2}\tilde\eta(q_1)\tilde\eta(q_2)
\left[Z_\la^{(2)}(\phi_0)(p_1^2+2q_1^2+q_1q_2+2q_1p_1)+U_\la^{(4)}(\phi_0)\right
]
\delta(p_1+p_2+q_1+q_2)\nonu
&+&O(\tilde\eta^3,k^4)\nonumber
\eea
Finally one computes the inverse of the operator 
$\la\cm_{p_1,p_2}+\Gamma^{(2)}_{p_1,p_2}$ and expands it in powers of 
$\tilde\eta$ and $k$. The trace over $p_1$ and $p_2$ needs the 
computations of terms like
\be
\mbox{Tr}\left\{(p_1q_1)(p_2q_2)F(p_1,p_2)\delta(p_1+p_2+q_1+q_2)\right\}=
\frac{q_1^2}{d}\delta(q_1+q_2)\int_p p^2F(p,-p)
\ee
and they lead to (\ref{evoluz}). The consistency with the 
equation for $U_\la^{(2)}$ is satisfied.

\section{Evolution equation in the external space}
The evolution for the blocking in the
momentum space is given by the Wegner-Houghton equation \eq{wheq}. 
In order to obtain its simplified version in the gradient expansion
one needs the trace of
the logarithm of $\Gamma^{(2)}_{p_1,p_2}$. This time the trace has to 
be computed for $|p_1|$ and $|p_2|$ in the shell $[k-\delta k,k]$. This
implies that $|p_1+q|$ has to be in the shell, as well, because 
$\Gamma^{(2)}_{p_1,p_2}$ contains $\delta(p_1+p_2+q)$.
This constrain implies the appearance of $\sqrt{q^2}$ in 
the equation for $k\partial_k Z_k(\phi)$. The terms containing $\sqrt{q^2}$ 
are non-local and spoil the the gradient expansion. 
$k\partial_k Z_k(\phi)$ is proportional to the second derivative
of the trace in \eq{wheq} with respect the momentum $q$ of the infrared
background field, $\tilde\phi$. The trace can be written as a momentum internal 
over
the shell $[k-\delta k,k]$. There are contributions from the
dependence on $q$ of the integrand and the limit of integration. 
It is worthwhile noting that the later contains all non-local terms.

There are two ways to rid the non-local contributions when the model is
solved by the loop expansion, i.e. by means of loop integrals for momenta 
$0\le p\le\Lambda$. One is to use lattice regularization where the
periodicity in the Brillouin zone cancel the $q$ dependence of
the domain of the integration. Another way to eliminate the non-local
terms is to remove the cutoff. Since the non-local contributions
represent surface terms they vanish as $\Lambda\to\infty$.

One may furthermore speculate that some of the non-local terms cancel
between the consecutive steps of the blocking $k\to k-\dk$
for a suitable choice of the cutoff function $f(\kappa)$
in the propagator $G^{-1}_k(p)=f(p/k)G^{-1}(p)$. 
Ignoring simply the non-local terms the identification of the
coefficients of the different powers of the gradient in the two
sides of \eq{wheq} leads to
\bea
k\partial_kU_k(\phi_0)&=&-\frac{\hbar\Omega_d k^d}{2(2\pi)^d}
\ln\left(\frac{Z_k(\phi_0)k^2+U_k^{(2)}(\phi_0)}
{Z_k(0)k^2+U_k^{(2)}(0)}\right)\\
k\partial_kZ_k(\phi_0)&=&-\frac{\hbar\Omega_d k^d}{2(2\pi)^d}\left(
\frac{Z_k^{(2)}(\phi_0)}{Z_k(\phi_0)k^2+U_k^{(2)}(\phi_0)}-2Z_k^{(1)}(\phi_0)
\frac{Z_k^{(1)}(\phi_0)k^2+U_k^{(3)}(\phi_0)}
{\left(Z_k(\phi_0)k^2+U_k^{(2)}(\phi_0)\right)^2}\right.\nonumber\\
&-&\frac{k^2}{d}\frac{\left(Z_k^{(1)}(\phi_0)\right)^2}
{\left(Z_k(\phi_0)k^2+U_k^{(2)}(\phi_0)\right)^2}
+\frac{4k^2}{d}Z_k(\phi_0)Z_k^{(1)}(\phi_0)
\frac{Z_k^{(1)}(\phi_0)k^2+U_k^{(3)}(\phi_0)}
{\left(Z_k(\phi_0)k^2+U_k^{(2)}(\phi_0)\right)^3}\nonumber\\
&+&\left.Z_k(\phi_0)
\frac{\left(Z_k^{(1)}(\phi_0)k^2+U_k^{(3)}(\phi_0)\right)^2}
{\left(Z_k(\phi_0)k^2+U_k^{(2)}(\phi_0)\right)^3}
-\frac{4k^2}{d}Z_k^2(\phi_0)
\frac{\left(Z_k^{(1)}(\phi_0)k^2+U_k^{(3)}(\phi_0)\right)^2}
{\left(Z_k(\phi_0)k^2+U_k^{(2)}(\phi_0)\right)^4}\right).
\nonumber
\eea
When $k^2\gg U^{(2)}_k(\phi)$ this gives \eq{equazk} in dimension $d=4$.

\end{appendix}



\begin{references}
\bibitem{wilsrg}K. Wilson and J. Kogut, \Journal{\PREPC}{12}{75}{1974};
K. Wilson, \Journal{\RMP}{47}{773}{1975}. 
\bibitem{wh} F. J. Wegner, A. Houghton, \Journal{\PRA}{8}{40}{1973}. 
\bibitem{polc} J. Polchinski, \Journal{\NPB}{231}{269}{1984};
\bibitem{wett} C. Wetterich, \Journal{\PLB}{301}{90}{1993}; 
M. Reuter, C. Wetterich, \Journal{\NPB}{391}{147}{1993}.
\bibitem{morr} T. Morris, \Journal{\IJMPA}{9}{2411}{1994};
\bibitem{ellwang} U. Ellwanger, \Journal{\PLB}{335}{364}{1994}.
\bibitem{tetra} N. Tetradis, D.F. Litim, \Journal{\NPB}{464}{492}{1996};
J. Adams, J. Berges, S. Bornholdt, F. Freire, N. Tetradis, C. Wetterich,
\Journal{\MPLA}{10}{2367}{1995}.
\bibitem{cowe} S. Coleman, E. Weinberg, \Journal{\PRD}{7}{1888}{1973}.
\bibitem{casy} C. G. Callan, \Journal{\PRD}{2}{1541}{1970};
K. Symanzik, \Journal{\CMP}{18}{227}{1970}.
\bibitem{glob} J. Alexandre, V. Branchina, J. Polonyi, 
\Journal{\PRD}{58}{16002}{1998}.
\bibitem{devgrad} T. Morris, \Journal{\IJMPB}{12}{1343}{1998};
\Journal{\PLB}{334}{355}{1994}; \Journal{\PLB}{329}{241}{1994};
T. Morris, M. Turner, \Journal{\NPB}{509}{637}{1998}.
\bibitem{tree} J. Alexandre, V. Branchina, J. Polonyi, 
{\em Instability Induced Renormalization},
to be published {\em Phys. Lett.} {\bf B};
J. Alexandre, J. Polonyi, {\em Tree-level Renormalization}, 
submitted to {\it Phys. Rev.}.
\bibitem{schd} R. D. Ball, P. E. Haagensen, J. I. Latorre, E. Moreno,
\Journal{\PLB}{347}{80}{1995}.
\bibitem{fraser} C.M Fraser, \Journal{\ZPC}{28}{101}{1985} 
\bibitem{simi} M. Simionato, {\em Gauge Consistent Wilson Renormalization
Group I., II.} hep-th/9809004, 810117.
\end{references}
\end{document}